\magnification=\magstep1  
\vsize=8.5truein
\hsize=6.3truein
\baselineskip=18truept
\parskip=4truept
\vskip 18pt
\vskip 18pt
\def\today{\ifcase\month\or January\or February\or
March\or April\or May\or June\or July\or
August\or September\or October\or November\or
December\fi
\space\number\day, \number\year}
\centerline{\bf SIMULATED BILAYER from FLOPPY to HOLE and TOPOLOGICAL }
\centerline{\bf FLUCTUATIONS. }
\medskip
\centerline{by }
\centerline{J. Stecki }
\medskip   
\centerline{ Institute of Physical Chemistry, Polish Academy of Sciences,}
\centerline{ ul. Kasprzaka 44/52, 01-224 Warszawa, Poland}
\bigskip
\centerline{\today} 
\vskip 60pt
\centerline{Abstract. }
\vskip 20pt
Simulations  of a bilayer in a liquid solvent in the full range of areas
from the floppy to the extended state also show the hole/tunnel spontanous 
formation. Whether this change of topology is preceded by topological 
fluctuations of the Gaussian curvature $K(x,y)$ is an open question.
 Probability distributions  of Gaussian curvature are 
transformed into those of the ordered inverse radii;
these new shapes show a gradual change along the isotherm.
The model of Max-Planck group with 5 beads is used, amended to 
represent amphiphiles in the spirit of the "coarse-grained" or "Martini"
force field.\vfill\eject 

{\bf Introduction.}

Simulations  of  bilayers have greatly extended the knowledge about and 
the understanding of these prototypes of membranes. Whereas most work
was concentrated on the simulation of tensionless bilayers, the wider
range of areas was examined as well, beginning with a small interval 
in the pioneering work of this kind[1], up to the entire range of 
the existence of a stable bilayer[2-7].  The simplest setup is the 
liquid system at constant temperature $T$, volume $V$, and constant 
$N$, the number of amphiphilic molecules, and $N_s$, the number of 
solvent molecules. With change of shape the area $A$ varies and the 
lateral tension $\Gamma$ with it; the  "bilayer isotherm"
$\Gamma (A)$ results[2-7]. Its unusual and interesting features are
briefly recalled in Section I. One may inquire as to the effect of 
temperature, but with one exception[7] and another[8] this 
has not been examined. We show some results elsewhere. 
With the increase in the area $A$ at constant temperature, the 
stretched bilayer eventually gives way but it does so in an interesting 
manner; it does not break up in droplets or micelles but recedes to keep 
its surface density and creates a hole $-$ a tunnel $-$ through which 
the disjoined portions of solvent may connect. Thus the hole
 is filled with the solvent. The process of such spontaneous formation,
without any foreign body attempting to pass through the membrane, has
not been studied much by simulation[9]; we examine its appearing on
the bilayer isotherm, in relation 
to the change of Gaussian curvature $K$. Normally this quantity is of 
no interest as it is null and constant, but with the hole formation 
the genus changes (from the value $p=-2$ appropriate for system with 
periodic boundary conditions) and therefore $K$ changes. Section II
is devoted to the Gaussian curvature. An interesting question 
is whether the sudden creation of hole is preceded by topological 
fluctuations, fluctuations in $K$.
\bigskip    

{\bf Section I. The bilayer isotherm and the model used.}

Fig.1 shows a bilayer isotherm, the lateral tension $\Gamma$,  
plotted against the specific area per surfactant molecule $a = 2A/N$. The 
main portion of the graph shows the extended bilayer, (EX), where $\Gamma$ is 
positive and increasing. Compressed bilayer reaches the tensionless state
(which we call the "tensionless state nr.1"); at this area $a_1$ $\Gamma$ 
vanishes. On further compression of the area the bilayer undergoes a
transition to the "floppy" state (FL). This is the small area of the graph 
where $\Gamma$ is negative. Such transitions and the existence
of the floppy state are known from  experimental studies of vesicles[10]
and a counterpart is found here. The surprising feature is the 
{\it constancy} of the {\it negative} $\Gamma$. All this can be understood
and qualitatively described by the very simplest theory of undulations as
put forward by Helfrich[11]. We have also shown that the scaling with size is
predicted by that theory and confirmed the prediction by our simulations of 4 
systems of 4 sizes from $N=450$ to $N=4000$[4]. Undulations of bilayers 
have been much studied, for example in[2-12], and the corresponding power spectrum of fluctuations,
often under the name of the structure factor, $S(q)$, has been determined
repeatedly. $S$ is expected to diverge on vanishing $q$, $q\to 0^+$,
as $1/q^2$  for capillary waves and as $1/q^4$ for tensionless bilayer
(in the "tensionless state nr 2"). Now, upon the transition to the floppy state
the asymptote $q=0$ is replaced by an asymptote at a positive value of 
$q$, $q=q_{as}>0$  and $S_{floppy}(q)$ diverges as $1/(q^2-q_{as}^2)$. The region
$ q \in (0,q_{as})$ is not accessible. It can be shown[5] that such shift
of the $q$-asymptote is also predicted by the Helfrich undulation theory.
The latter is a heuristic adaptation of the capillary-wave theory but it 
has been put on a firm footing in a mathematical paper[12] which derives all 
of it from a microscopic bending hamiltonian, by a coarse-graining procedure.
It also follows from that theoretical description that the transition itself
is not a true phase transition but a crossover - despite the appearances.

On further compression of the already floppy bilayer there comes a point when
the bilayer disintegrates into what appears to be a structureless foam and
sometimes a "quad-layer". 

The picture described and seen in the left-hand side of Fig.1 is valid for
sufficiently large systems. For small bilayers the EX portion of the 
isotherm continues  downwards to  large negative values of $\Gamma$ and 
then suddenly starts to increase strongly - so that the curve $\Gamma (a)$ 
has a spike - and there is no flat section of constant, or nearly constant
 $\Gamma$[4,5]. In our experience at least 1200 amphiphiles were necessary
 to produce the expacted floppy plateau.

The bilayer with the specific area increased above $a_1$ is in an extended state,
gently undulating,  with $\Gamma$ increasing less than linearly, with $a$; then
becomes  stretched and finally does not disintegrate but recedes and creates a 
hole. The hole is nearly cylindrical and is filled with the solvent. The first 
published picture[3] of such a hole already showed that the bilayer recedes
keeping its surface area; this was superseded by a more thorough 
investigation[9]. The lateral tension drops to much lower value.
The downward trend of $d\Gamma /d a$ in Fig.1  suggests that perhaps the 
hole is formed when $d\Gamma /d a = 0$ but on closer scrutiny the plot does 
not fully support that; the derivative seems to be still positive at 
metastable states. 
 
The details of the shape of a bilayer isotherm will depend on the model used
and on the details onf the intermolecular force field.
In Fig.1 we show two sets of points, supplemented by a few points (shown with 
circles and line) at constant pressure and (slightly) variable density. This is
meant to imply that it does not matter very much that we take isochoric 
data, at constant (overall) density. The main points (Diamonds) were obtained
with the additional repulsive force (ARF) mimicking the hydrophobic effect 
after Goetz et al.[1]; the smaller set of data (Stars) - without that 
"Deus-ex-machina" addition.
  
Fig.2 shows a similar isotherm but for molecules with longer 8-bead tails. Since 
the system is also bigger, the floppy flat portion of $\Gamma (a)$ is better 
developped and truly flat.

 The model we used in the past was introduced by the Max Planck group[1],
 as an extension of the very first modelling of amphiphilic molecules - as 
dimers in a solvent[13]. The liquid system is made of spherical particles, which 
are either solvent molecules or beads connected in linear chains  
of unbreakable (chemical) bonds to form  amphiphilic molecules. 
In such a molecule, one terminal bead represents the polar 
head and the remaining - the nonpolar hydrocarbon chain.
Here we use 5 beads ({\it i.e.} 1 for the head plus 4 for the tail) and exceptionally
 9 {\it i.e.} one for the head and 8 for the linear chain as tail.
The intermolecular energy is a sum of pair interactions, all of which are
cut and shifted 6-12 Lennard-Jones functions:
$$ u(r))=4\epsilon {\bigl [(\sigma /r)^{12} -(\sigma /r)^6 \bigr ]} \eqno(1)$$  
In general for 3 species (solvent "s", head "a", and tail-bead "t")
the symmetrical matrix of parameters requires 6 energies
$\epsilon_{\alpha\beta}$ and another requires 6 collision diameters 
$\sigma_{\alpha\beta}$. Practically in most if not all simulation work the 
collision diameters were taken equal for all pairs. Often  the
energy parameters were also made all equal[1,4,14] as well. 
Then, as one author noted[14] the molecule was not amphiphilic
any more. 

Also, an extra term was added to substitute for the absent
"hydrophobic force". It is well-known that a hydrocarbon molecule "prefers" 
its own environement to water and vice versa; it will experience an effective 
repulsion, in spite of all dispersion forces being attractive. 
Such an additional repulsive force (ARF) between the solvent "s" and the tail
 beads "t", has been added[1] and  assumed to vary as $1/r^{n}$ 
with $n=8$ or 10. 

To improve on that model, we note that
the essential feature of an amphiphilic molecule is the
permanent chemical bonding of two antagonistic groups - a polar 
(hydrophilic) head and the hydrophobic (non-polar) tail, 
most often a hydrocarbon tail. 
Accordingly, we treat these two groups of particles differently.
The $\epsilon$'s are taken all equal within the non-polar group 
 $$ \epsilon\equiv \epsilon_{tt}=\epsilon_{at}=\epsilon_{st} , \eqno(2)  $$ 
and augmented for the "polar group":
$$\epsilon_{ss}=\epsilon_{sa}=\epsilon_{aa}=\omega\epsilon . \eqno(3) $$ 
With $\omega > 1$ the polar beads "s" and "a" are 
endowed with a stronger mutual attraction.

In this simple manner the model is improved and the molecule cannot be said
"not to be amphiphilic any more".
In most of our work we have used the value $\omega = 2.$ .
For all intermolecular forces all $\sigma$'s are taken equal.  
The cutoff at distance $r = 2.5\sigma$ is common to the polar group. 
Next one manipulates further the potentials after Toxvaerd[15].
All potentials must be shifted if cut, so as to ensure continuity of forces.
The {\it s-t} and {\it a-t} forces  are  made repulsive at all distances
by changing the cutoff to $r=\sigma$ (and therefore the shift to $+\epsilon$).
The {\it intra}molecular bonds
are confined to the small neighbourhood of the optimal bond distance by a 
potential well with infinite barriers[4,15].

Our model can be viewed as a simplified version of the "Martini force
field"[16], the latter also named a "coarse grained" model[17].

The Molecular Dynamics simulations were done in the canonical ensemble {\it i.e.}  
at constant $T,V,N$ with Nose-Hoover thermostat. 
\bigskip

{\bf Section II. The hole formation and the topological fluctuations. }

Thw spontaneous hole formation in the stretched bilayer is a change 
of topology. It will affect the Gassian curvature $K$. Whether this 
sudden change of the surface (and of its genus) is preceded by 
fluctuations, e.g. by fluctuations of $K$, is not known.
It is accepted that  approaching even a sharp
 first-order order transition, "the system and its molecules exercise 
themselves" [18] in preparation to the transition.
We have examined the behaviour of $K$ along the bilayer isotherm reported
above. If the Fourier coefficients have been determined, no matter how,
one can construct the surface $h(x,y)$ and from the definition of $K$ to 
obtain $K(x,y)$. Total, or overall, $K_t$ is then
$$ K_t = \int_0^L \int_0^L dx dy K(x,y)               \eqno(4) $$
If a surface is given by the smooth function $h(x,y)$ in the Monge gauge then
$$ K(x,y) \equiv (h_{xx}*h_{yy} - h_{xy}^2)/D^4 \eqno(5) $$
where 
$$ D\equiv \sqrt{1 + h_x^2 +h_y^2}  \eqno(6) $$
in the usual notation of first and second derivatives of $h$.
From the Fourier analysis of the simulation data {\it i.e.} positions of heads,
$h$ is given as a finite sum
$$ h(x,y)= a_0/2 +\sum a_n \cos(q_n R_n) + b_n \sin(q_n R_n) \eqno(7) $$
where $q_n$ is a vector $(2\pi/L)(n_x,n_y)$ and $R=(x,y)$.
The $q$-vectors are usually chosen to fill a square centered at $(0,0)$ and
a half of it is used in summation, coresponding to the  square
area $0 < x <L$, $0 < y < L$. The other curvature, the mean curvature $H$ is 
also obtained from $h(x,y)$ by standard expression, as always with $D\not= 1$.

Both were obtained from $h(x,y)$ at a set of arbitrary values $(x_j,y_j)$, 
chosen as a square grid sufficiently fine for trapezoidal integration. 
The latter  produced the overall total $K_t$ and total $H^2$,
 their dispersions, and a large set of values of $K$ at $(x_j,y_j)$ to be
used for the construction of histogram. These computations were done
along the entire simulation run and the sets of data obtained were 
used to construct histograms $P(K)dK$ and $P(H^2)d(H^2)$. 

The histograms for $H^2$ are shown in Figure 3.  The mean curvature
squared $H^2$ decays monotonously attaining a rate not too far from an 
exponential decay in the latest stage of the decay, as can be seen from 
Figure 3 in semilogarithmic scale. The four points on the isotherm are
specified in full detail below;
the two upper curves for $L=53$ and $L=52$ are indistinguishable; that for 46
lies slightly higher than that for 38. The next Figure, Fig.4,  shows a portion of 
the same histograms, near $H^2=0$. The sequence is the same along the entire
histograms (the curves never cross) and the Figure also shows the steep 
fall very close to $H^2=0$.

The histograms for $K$ are more interesting.  
 $P(K)$ appears as  a Dirac delta-function, $\delta(0)$,
 in agreement with the expectation that for a two-dimensional surface
with periodic boundary condition the Euler genus equals -2 and therefore 
$K=0$. Fig.5 shows the remarkable distributions of $K$, in semilogarithmic 
scale to  reveal the broadening of the $\delta$-functions. 

The nature of the peak is examined in Fig.6 by using a magnified scale; remarkably
the discontinuity at $K=0$ is approached with a finite slope and with an
apparent lack of symmetry about $K=0$. These two features: discontinuity with 
finite slope and 
no symmetry, are seen again in Fig.7. It shows the plots of the inverse 
$1/P(K)$, very convenient for the estimation of the slopes and extrapolation
to the point of discontinuity. It is not clear if the extrapolation from the 
left produces the same value as the extrapolation from the right, in view
of different slopes and second derivatives. The mathematical representation 
is an open question; a gaussian distribution is certainly excluded as 
there is no inflexion in sight. An inverse power is not acceptable either,
as the limited plot of Fig.7 already shows.

$K$ was computed often, but not for flat bilayers; rather for those intricate 
surfaces formed in microemulsions and/or  general  minimal surfaces(cf.[19,20]).
 In these  nonplanar intricate shapes
the Euler index is changed and $K$ is a most important quantity.
There are no known expression for its distribution, even for one of 
the simplest minimal surfaces[19]. The shapes of those histograms, 
numerically computed[19], were very similar to those seen in 
our Figure 5.

With the  discontinuity at $K=0$ {\it i.e.} with the part with $K<0$
differing from the part with $K>0$ the task becomes even more complex.
But this will be unnecessary as will become clear.  
 
 All histograms  as numerical data were  obtained for 
different states of the bilayer, {\it i.e.} for different areas along the isotherm,
Four points were chosen: (1) almost at the tensionless state, (2) in the middle
of the isotherm, (3) near the breaking point of stretched bilayer and 
(4) at the breaking point where the hole-tunnel is formed.
The details are as follows: $T=1.35, N=1800, 4$ beads, $\rho =0.89$ for 
 l points; and then (1) $L=38.25,~ a=1.62525; (2)L=46.0,~ a=2.3511; (3)L=52.0,~
  a=3.004;$ and $(4) L=53.0,~  a=3.1211$. The last point is the point 
of hole formation; as mentioned earlier, there is hysteresis.
There are metastable states when the hole is not formed for a very long time.
for example, starting from the system at $L=52.$, after preliminary 
equilibration at $L=53.$, running with $L=53.$, we find no hole until 
~0.8E6 timesteps. Then expansion to
$L=54.$ of the system with newly formed hole, stabilized the tunnel structure,
which on compression back to $L=53.$ did not disappear.  
 The earlier part of the data for $L=53.$,{\it i.e.} before the hole was starting to 
form. were included as (4) and is codenamed as "53h". As is apparent from Fig.1
and 2, the hole formation makes the lateral tension $\Gamma$ to fall to very
different and low values[9]. 
 
For examining the approach to state (4) along the bilayer isotherm. {\it
i.e.} in the
sequence $1\to 2\to 3\to 4$  the plot of inverses $1/P(K)$ appears
the best; the peak broadens as seen from Fig.7 but there is no regular
pattern. 

The lack of symmetry about $K=0$ is a disturbing feature. We turn therefore
to the origin of these curvatures. 

The primary quantities are the {\it principal radii} of the surface $h(x,y)$
at that point, $R_1(x,y)$ and $R_2(x,y)$. The principal curvatures are 
defined as $c_1=1/R_1$ and $c_2=1/R_2$.
The principal radii are obtained as roots of a quadratic equation, usually 
written in terms of the $c$'s
$$ a*x^2 + b*x + c == a*(x - c_1)(x - c_2) = 0 .  \eqno(8) $$
As it turns out, $H$ and $K$ are related to the roots as
$$ H = (1/2)(c_1+c_2)  \eqno(9) $$
$$ K = c_1 c_2        \eqno(10) $$
Therefore $H$ is simply related to one coefficient in the quadratic
equation and $K$ to the other. Simple algebra produces the solution,
the roots, in terms of $H$ and $K$, are
$$ c_1= H - \sqrt{H^2-K} \eqno(11) $$  
$$ c_2= H + \sqrt{H^2-K} \eqno(12) $$  
In these relations of differential geometry  $H$ and $K$ are just abbreviations
for the combinations  of the first and second derivatives of $h(x,y)$. Also
$$ R_1 = (1/c_1)= c_2/K  \eqno(13)  $$
$$ R_2 = (1/c_2)= c_1/K  \eqno(14)  $$

A dump of the $R$'s and $c$'s  followed by the construction of their 
histograms, revealed interesting features. First, one radius is always very 
large in absolute value as compared to the other. This is due to near-planarity
of these surfaces. 
If $\vert R_2\vert$ is O(1), $\mid R_1\mid$ is 10000 or 
more. However, the signs vary; though $c_1 < c_2$ always, not necessarily 
$\mid c_1\mid < \mid c_2\mid $ -  so the smaller root
may be the big one in absolute value and conversely. 

The output roots were therefore ordered according to their absolute values:
if $\mid c_1\mid < \mid c_2\mid $ then  $c_b= c_1$, $c_s=c_2$,
and conversely, 
if $\mid c_1\mid > \mid c_2\mid $ then  $c_b= c_2$, $c_s=c_1$.
The convention is to use the subscript "s" for the {\it radius} smaller of 
the two, $\mid R_b\mid > \mid R_s\mid $ with $R_s=1/c_s$, $R_b=1/c_b$
and therefore for the larger of the two $c$'s.
It turns out that
which root will produce $R_b$ and which $R_s$, is decided by the sign of $H$,
the mean curvature. Simply comparing the squares: $ c_2^2>c_1^2$ if $H<0$ 
and conversely. And the value is 
$$ (R_s)^{-1} = c_s =(\mid H\mid + \sqrt{H^2-K})\times H/\mid H\mid .  \eqno(15)  $$
 The histograms must be now transformed accordingly.
First, histograms of unsorted roots $c_1$ and $c_2$ are shown in Fig.8, for
one point (2) ($L=46$). A peculiar symmetry appears in that the branch for
negative $K$ is the mirror image of the other histogram for positive $K$:
$$ P(c_1) = P(-c_2)  \eqno(16)  $$
The test for symmetry is better shown just for the positive semiaxis and 
with the inverses $1/P(c_1)$ and $1/P(c_2)$, as in Fig.9.

Now the histograms for sorted $c_s$ and $c_b$ take a very different look. 
Fig.10 shows $P(c_b)$ and $P(c_s)$, again for clarity for one state of the 
bilayer, namely  (2) with $L=46$.
The larger radius $R_b$ corresponds to $c_b$ and $P(c_b)dc_b$ has a
spike at $K=0$; the smaller radius is $R_s=1/c_s$ and $P(c_s)$ vanishes at
$c_s=0$ and has a maximum. There is perfect symmetry now between the negative
branch and the positive branch {\it of the same histogram}. $ P(c_s)$ starts
linearly from $P(0)=0)$ with the slope (1/4), goes through a maximum and then
falls quasi-exponentially; in the  final stage $P\sim const*\exp (-a*x)$. But 
as a fitting equation  has very limited significance as it is only valid  at 
very large values of the argument.  Fig.11 shows the pair from Fig.10 now with
semilogarithmic scale. It is seen that $P(c_s)$ decays almost exponentially,
as $\exp(-\alpha x)$ but with $\alpha = \alpha (x) $ drifting along the curve to
reach its final value at large values of the argument.  

The transformations to $R_s$ and $R_b$  substantiate and explain in simple terms
the curious symmetries seen in Figure 8 but above all remove the lack of 
symmetries in histograms of $K$. The essential point is that it was not
enough to use the principal curvatures $c_1$ and $c_2$; it is only the 
curvatures sorted according to their absolute values that had symmetrical
histograms. It is physically convincing that the irrelevant big principal 
radius $R_b$ and its curvature $c_b $, reflecting the overall near-planarity 
of the surface, were removed. The other and small radius $R_s$ and its
curvature $c_s$ are the relevant quantitities.

Now we compare $P(c_s)$ for all state points (1)-(4);  at last a
pattern of regular change appears. As the area per surfactant molecule 
increases, Fig.12 shows the sharpening of the distribution, a very small 
insignificant shift of the maximum, and constant initial slope at zero.
The decay portion of $P$ also changes regularly {\it i.e.} the decay is faster 
as the distribution gets sharper with the increase in area. 

From the histogram one obtains the average $<c_s> $ and its dispersion
$ \langle c_s^2\rangle -\langle c_s\rangle ^2$; these are shown for all 4 state-points in Fig.13.
Just as the histograms $P(c_s)dc_s$ there is smooth variation except for the
additional point at L=38; this is a state point with ARF.

In a related result, the variance of bare $K$ is shown in Fig.14.
Taken as $\langle KK\rangle - \langle K\rangle\langle K\rangle$, it shows good regular tends,
unlike all untransformed histograms. Because  $K=c_1c_2$, $K^2$ will be
$c_1^2 c_2^2 = c_s^2 c_b^2 $ and apparently the regular variation 
of $c_s^2$ is not spoiled much by the presence of $1/R_{big}^2$.
Note that $<K>$ may be slightly off its true value of nil; we take $D\not= 1$
so that $K(x,y)$ does not vanish identically as it does in the 
small gradient aproximation $D=1$. 

As mentioned earlier, it would be an unnecessary and tedious exercise to try to find a mathematicla representation fot the listograms of $K$. Also, even for
one of the simplest minimal surfaces, this is not known[19]. Now, for
the symmetric $P(c_s)dc_s$, it appears that a function
$$ P(x)  \sim  x \exp[-\alpha  x^n]~~~  (x \ge 0)   \eqno(17)$$
(with a mirror image for $x\le 0$) is not too far away, esp. for portions 
away from the maximmum of $P$. However, this attempt was not fully successful,
as different least-square values for $n$ were obtained: $n=0.668,~ n=1.~$ ,
and $ n=1.12, 1.25$. Trying to impose $n=1$ for all was clearly not correct.
And the maximum was not perfectly reproduced. Therefore a satisfactory
 mathematical 
representation, even for $P(c_s)$, was not found.

These results have some further consequences of minor nature.
The linear combination $c_1 + c_2$ or $c_1 - c_2$   
will be often swamped by the root bigger in absolute value and therefore might
not be very useful.
We note that  the quantity $H^2-K = (1/4)(c_1-c_2)^2 $ is always
positive, so that the quadratic equation always has real roots and never 
complex roots.

  As the histograms of $c_s$ were shown above, those for the radii,
  $P(R_s)dR_s$,  are not shown 
because their plots do not bring anything new.

\bigskip

{\bf  Summary and Discussion.}

We have found a new transformation of the curvatures $H$ and $K$, the mean and
the Gaussian, to quantities with exactly symmetric (about zero) histograms and
varying smoothly along the bilayer isotherm. 

The numerical results were shown for 4 selected state points along the 
bilayer isotherm, one near the tensionless state no.1,  one in the middle
of the isotherm, one almost at the breaking point of hole formation, and one
(metastable state) exactly at the hole formation. 
The data - positions of heads on both monolayers - were collected from 
simulation runs at $T,V,N$, one run producing e.g. 7600 data sets.

The untransformed histograms - probability distributions -  $P(H^2)d(H^2)$ and
\hfill\break  $P(K)dK$,  were examined and their plots are shown in the 
Figures. Their features
led to reexamination of the derivation of these curvatures, as it is given
 in standard differential geometry; a new transformation resulted. 
In it, the principal
radii are sorted according to their magnitude irrespectively of sign. 
This ordering is crucial; it is not sufficient to introduce principal
curvatures $c_1$ and $c_2$.

These new objects, $c_s=1/R_s$ and $c_b=1/R_b$, have 
symmetrical probability distributions which also vary  smoothly
along the isotherm.

The resulting pattern of variation along the isotherm, can be interpreted 
by recalling that along the isotherm $\Gamma (a)$, 
with the increase of the specific area $a$ the bilayer is 
gradually stretched. Therefore predictably stretching impairs strong fluctuations
of the monolayers and of the bilayer as a whole. Thus the distributions shown
in Fig.12 have smaller probability of large argument for the bilayer more 
strongly stretched. Similarly the dispersions fall with $a$, as shown by 
Fig.13 and Fig.14.

There is no premonition of the breaking of the bilayer surface by creation
of a hole - it appears to be a catastrophe without warning.
As pointed out in the introduction, it is  quite a  special phenomenon, when 
the bilayer spontaneously recedes, trying to keep its structure and surface 
density, and makes room for a tunnel, the latter immediately filled with 
solvent molecules. 

\bigskip
 
\noindent{\bf Acknowledgements.}

The author is greatly indebted to Dr John Nagle (Pittsburgh) for a discussion, 
but particularly to Dr O.Edholm (Stockholm) for several extensive discussions which led
the author to undertake this particular investigation. He thanks the Institute of Physical
Chemistry of the Pol. Acad. Sci. for their continuing support.

\bigskip
\bigskip

\centerline{\bf References       }

\item {[1]} R. Goetz and R. Lipowsky, J. Chem. Phys. 108, 7397 (1998).
\item {[2]} J. Stecki, Int. J. Thermophys. 22, 175(2001).
\item {[3]} J. Stecki, J. Chem. Phys. 120, 3508 (2004).
\item {[4]} J. Stecki, J. Chem. Phys. 122,111102(2005).
\item {[5]} J. Stecki, J. Chem. Phys. 125, 154902 (2006);arXiv.0412248\hfill\break 
               available as of Dec.10,2004.
               see also J.Phys.Chem. B112(14),4246(2008).
\item {[6]} W. K. den Otter, J. Chem. Phys. 123, 214906 (2005).
\item {[7]} H. Noguchi and G. Gompper, Phys. Rev.E73, 021903 (2006).
\item {[8]} E. Lindahl and O.Edholm, J. Chem. Phys.113,3882(2000)
\item {[9]} see however I. R. Cooke and M.Deserno, J. Chem. Phys.123,224710(2005).
\item {[10]} E. Evans and W. Rawicz, Phys. Rev. Lett. 64,2094(1990).
\item {[11]} W. Helfrich and R. M. Servuss, Nuovo Cimento 3D, 137 (1984);
             see also W.  Helfrich, in {\it  Les Houches, Session XLVIII, 1988,
             Liquids at Interfaces} (Elsevier, New York, 1989). 
\item {[12]}  A.Adjari, J.-B. Fournier, and L.Peliti, Phys. Rev. Lett.86,4970(2001).
\item {[13]} B. Smit Phys. Rev. A 37, 3431 (1988)
\item {[14]} A. Imparato, J. C. Shilcock, and R. Lipowsky, Eur. Phys. J. E11, 21 (2003)
\item {[15]} e.g. Paz Padilla and Soeren Toxvaerd, J. Chem. Phys.106,2342(1997).
\item {[16]} S.J.Marrinck, A.H.de Vries, and A.E.Mark, J.Phys.Chem.B108,750(2003).
\item {[17]} E. G.Brandt, A.R.Brown, J.N.Sachs, J.F.Nagle, and O.Edholm, 
             Biophysical Journal 100, 2104 (2011) and Supplement.
\item {[18]} B. Widom, at the  Amsterdam IUPAP Conf. on Statistical Mechanics, 1973.
\item {[19]} U. S. Schwarz and G. Gompper,  Phys. Rev. Lett.85,1472(2000) where 
             references to related work can be found.
\item {[20]} R. Holyst and W.Gozdz, in Encyclopedia of Appied Physics, Wiley-VCH,
             Weiheim 1999, pp 146-160, where references to earlier work can be 
             found. 
 
\bigskip
\bigskip
\centerline{\bf Figure Captions. }
\medskip
\noindent A short Caption is put under each Figure.\hfill\break
In all Figures the scales are arbitrary with one exception of normalized
distribution in Fig.12. 

\medskip

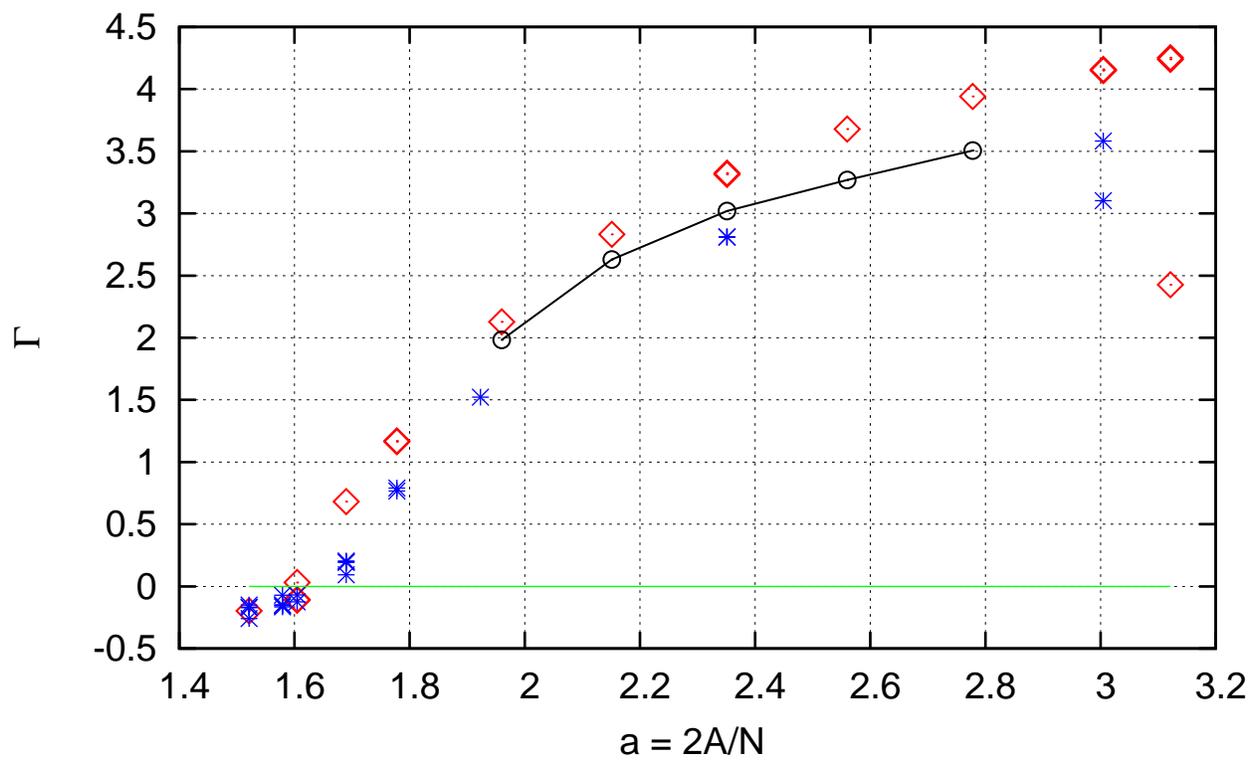
FIG.1. Bilayer isotherm T=1.35,4+1 beads, w/ extra force(Diamonds) and without
(Stars); Circles - $P$ = const.

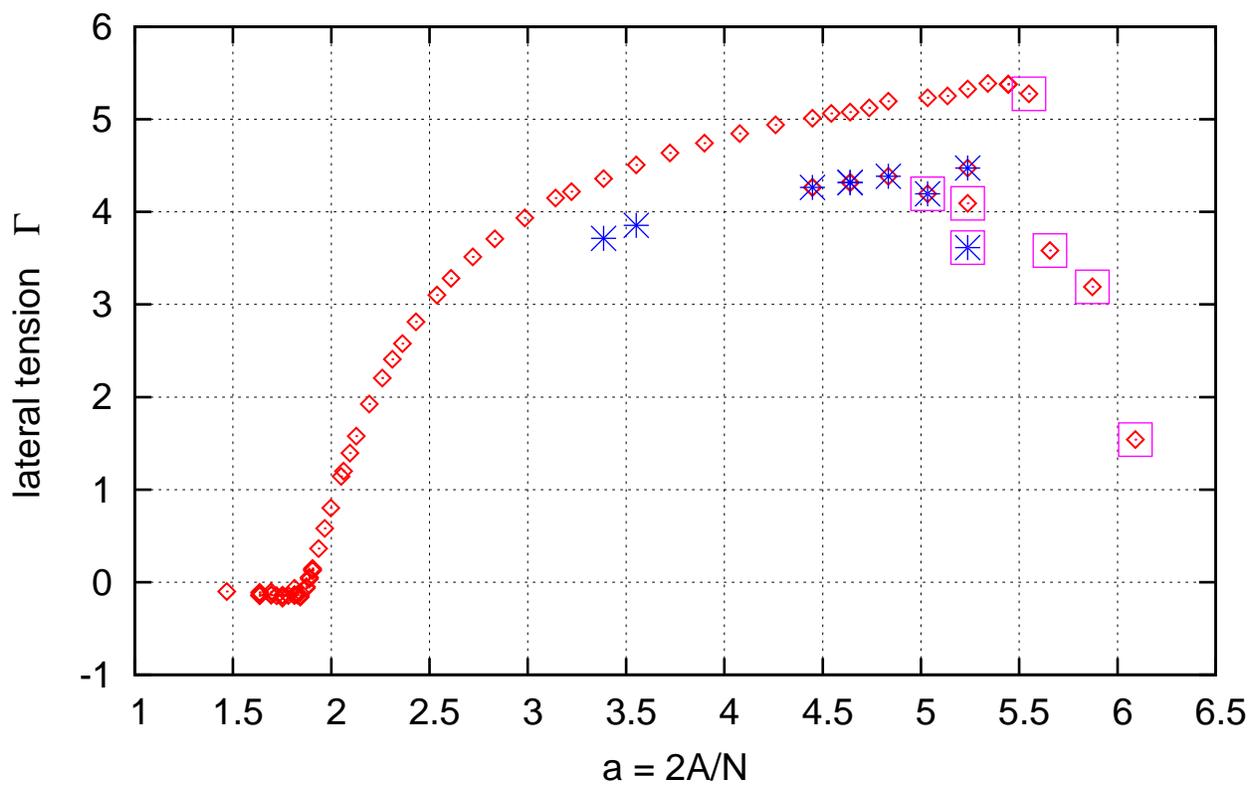
FIG.2. Isotherm T=1.35 with (Diamonds) and without(Stars) the extra-force AFR;
Boxes mark runs with holes. 8+1 beads.

\medskip

FIG.3. $P(H^2)d(H^2)$ for L=53(hole),52,46,38.

\medskip

FIG.4. $P(H^2)d(H^2)$ for L=53(hole),52,46,38. 

\medskip

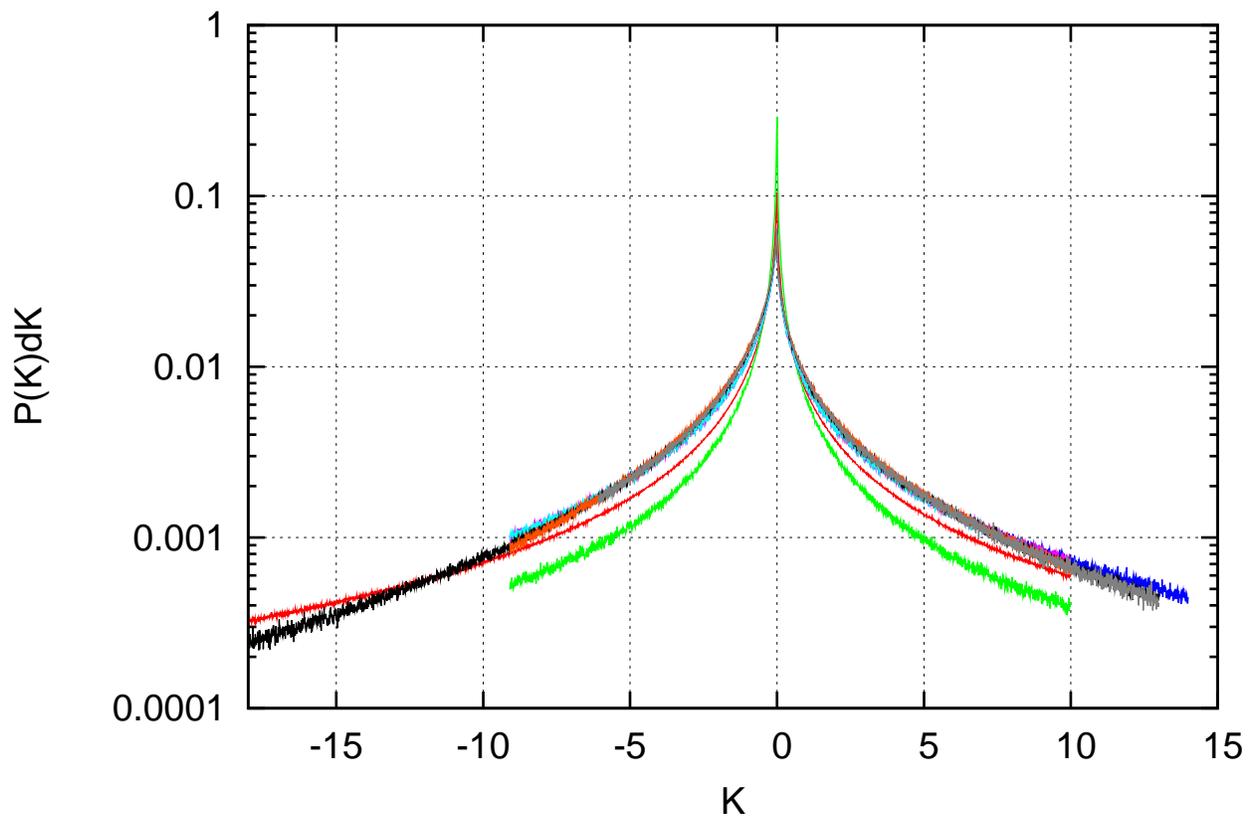
FIG.5. Histograms of Gaussian K for L=38,46,52,53,53h. 

\medskip

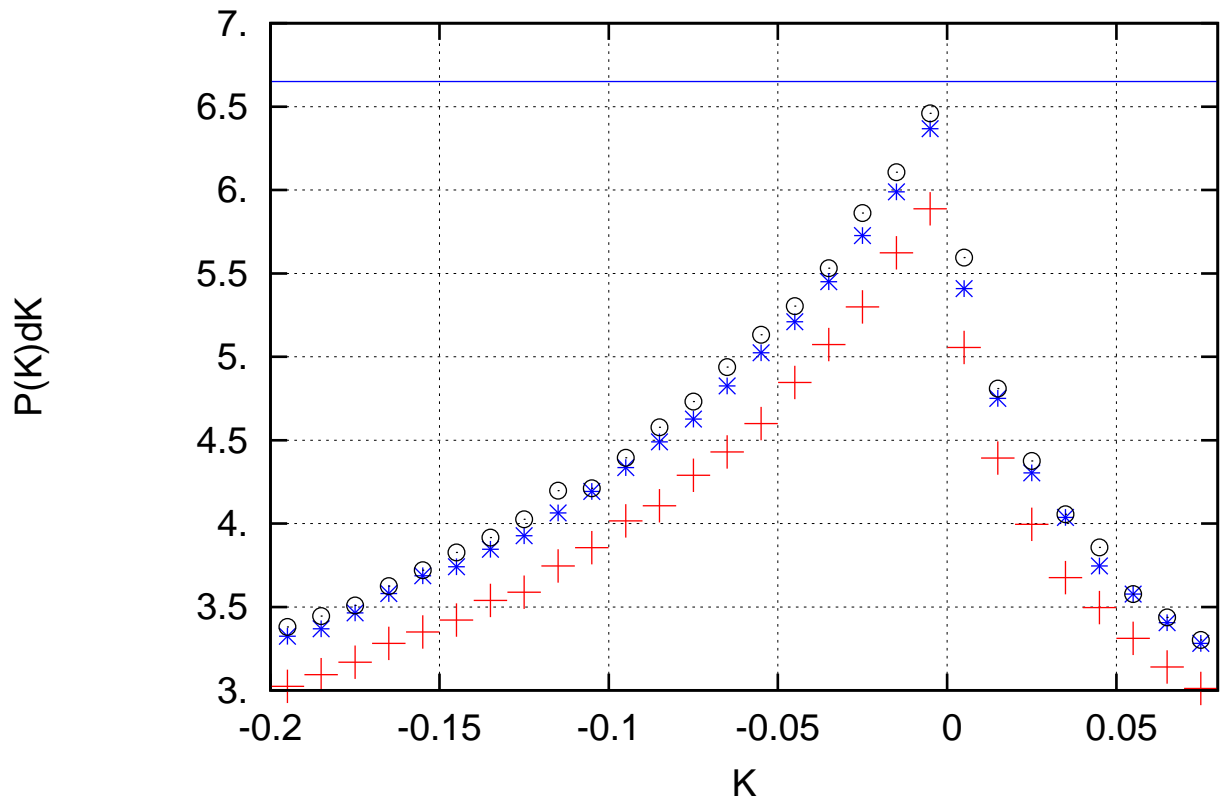
FIG.6. Selected peaks of Gaussian K; L=46(+),52(Stars),53h(Circles) and estd. 
maximum -line.

\medskip

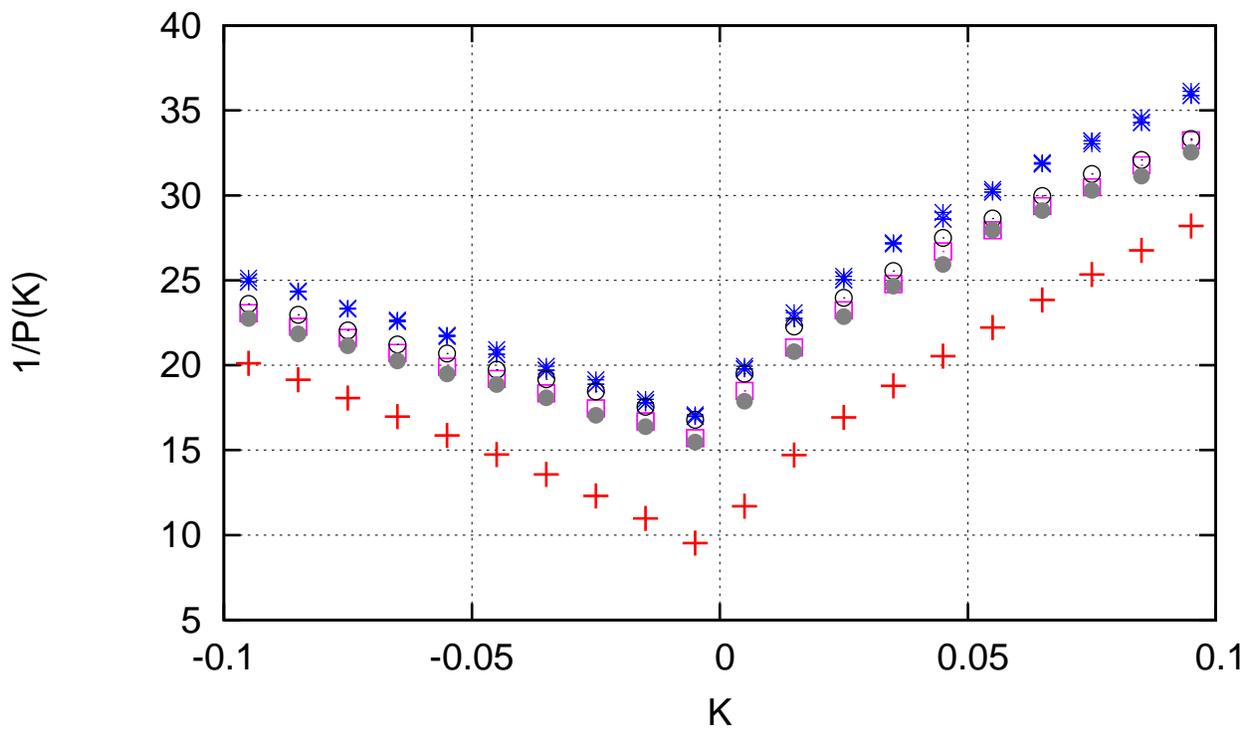
FIG.7. Inverse of $P(K)$ near peaks: L=38(+), 46(Stars), 53(Circles), 53h(Filled Circles).No pattern, no symmetry.

\medskip

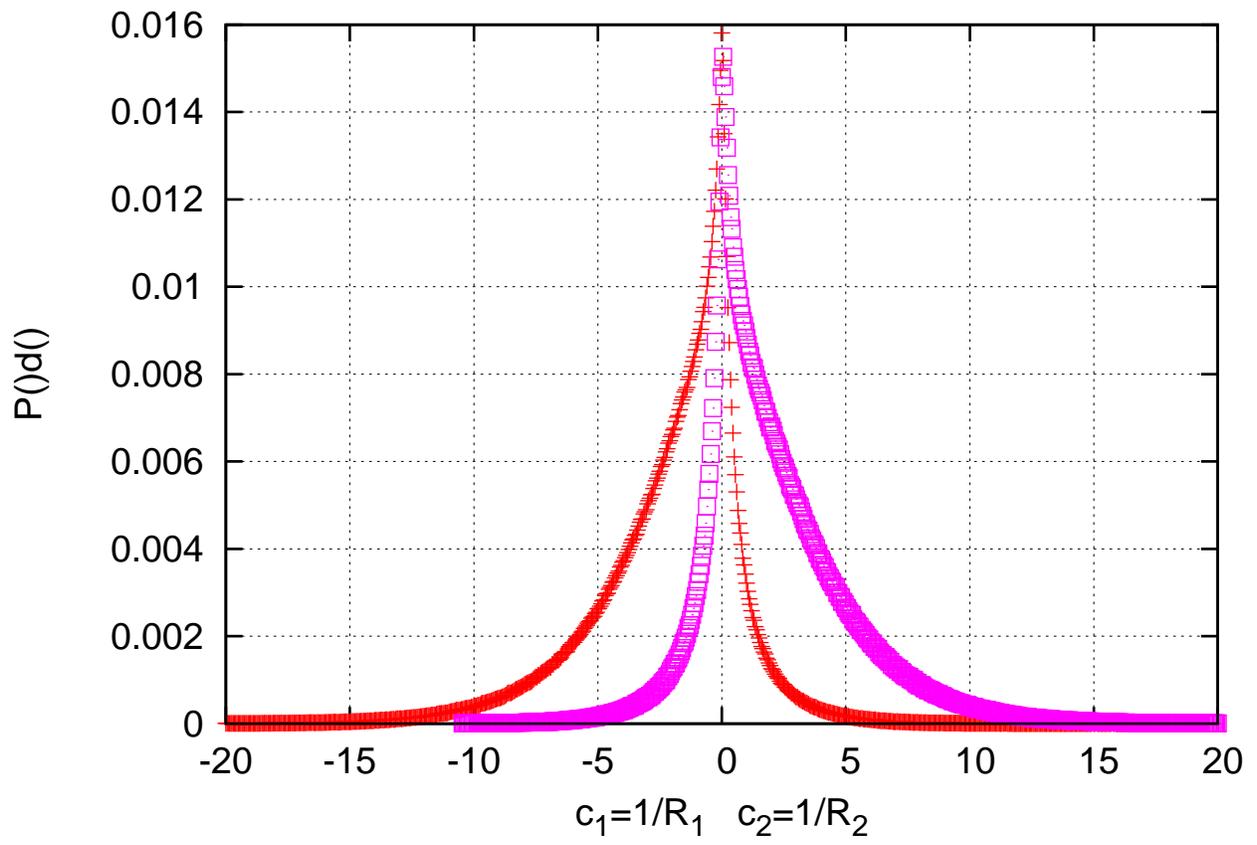
FIG.8. Principal curvatures; L=46.

\medskip

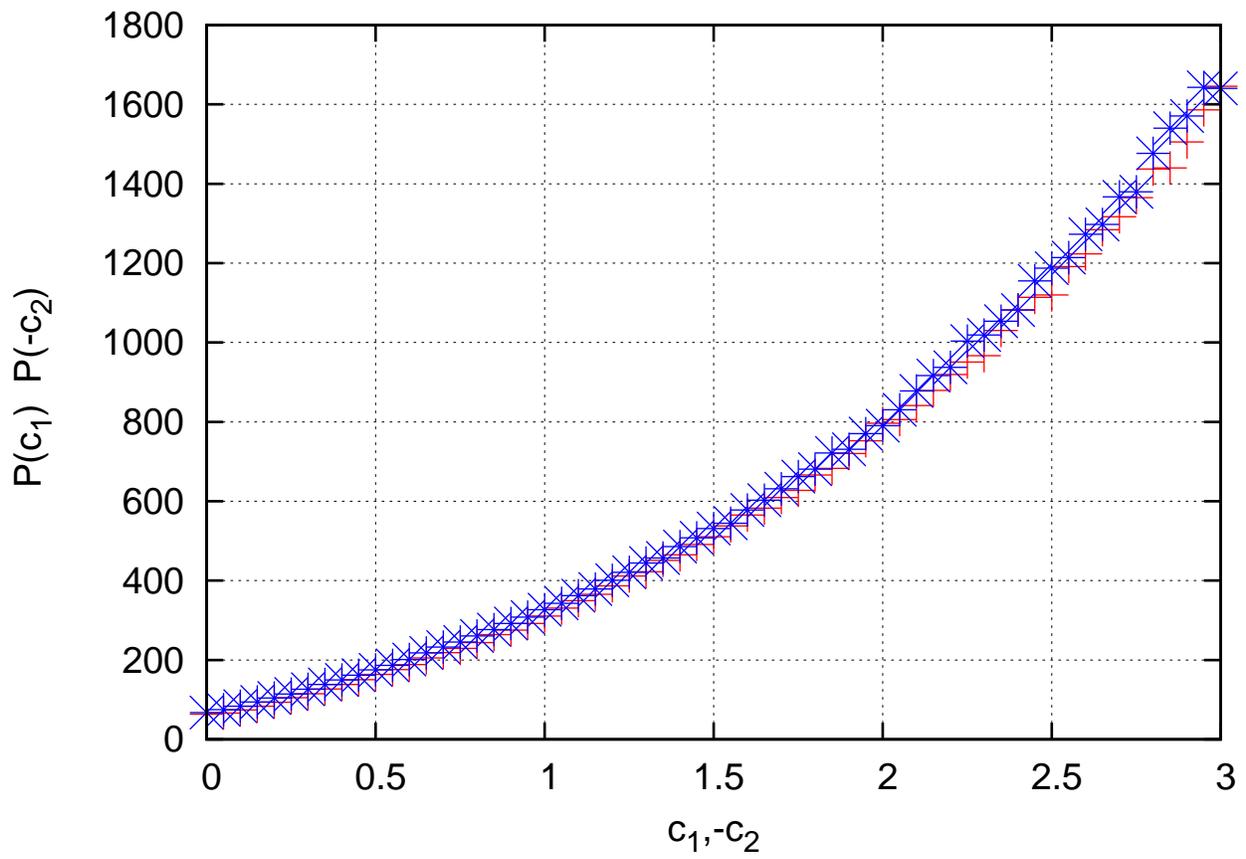
FIG.9. Symmetries for principal curvatures. Proof that $P(c_1)=P(-c_2)$.

FIG.10. Sorted roots (principal curvatures); the radius $R_b$ is bigger in absolute value; 
$c_b=1/R_b$ - Plus signs, $c_s=1/R_s$ - Stars.

\medskip

FIG.11. Sorted roots (principal curvatures); the radius $R_b$ is bigger in 
absolute value; $c_b=1/R_b$ - Plus signs, $c_s=1/R_s$ - Stars.

\medskip

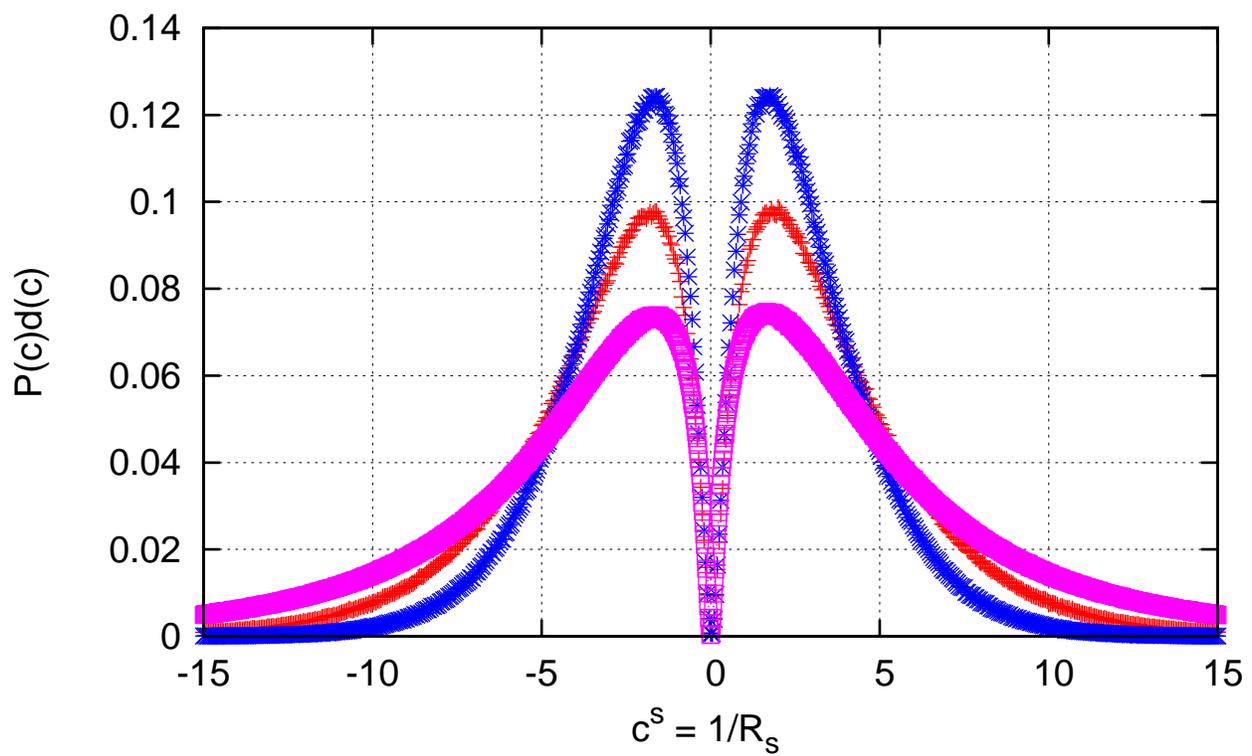
FIG.12. $P(c_s)$ normalized to (1/2), with $R_s$ being the smaller in absolute value;
$L=38$(Boxes), $L=46$(Plus signs), $L=53$(Stars).

\medskip

FIG.13. Curvature $c=1/R_{sm}$; $\langle c\rangle $(Diamonds),$\langle c^2\rangle $(Stars),
$\langle (c-\langle c\rangle )^2\rangle $ (Boxes). 

\medskip

FIG.14. Dispersions for H(Stars),H(sm.grad.)(Circles),K(Boxes).

\vfill\eject\end
\bye